\pdfoutput=1
\documentclass[twoside,a4paper,11pt]{sea10}
\usepackage{graphicx}
\usepackage{hyperref}
\usepackage{movie15}
\begin{document}
\pagenumbering{arabic}
\pagestyle{myheadings}
\thispagestyle{empty}
{\flushleft\includegraphics[width=\textwidth,bb=58 650 590 680]{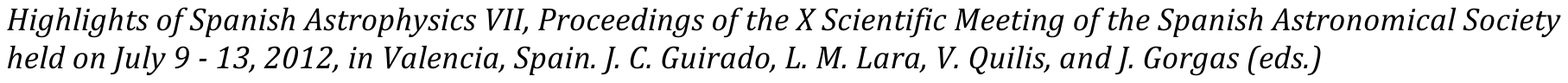}}
\vspace*{0.2cm}
\begin{flushleft}
{\bf {\LARGE
%
CARMENES. I. A radial-velocity survey for terrestrial planets in the habitable zones of M dwarfs. A historical overview.
%
}\\
\vspace*{1cm}
%
P.~J. Amado$^{1}$,
A. Quirrenbach$^{2}$,
I.~Ribas$^{4}$,
J. A. Caballero$^{3}$,
M.~A.~S\'anchez-Carrasco$^{1}$,
A.~Reiners$^{5}$,
W.~Seifert$^{2}$,
R.~Mundt$^{6}$,
H. Mandel$^{2}$,
and the CARMENES Consortium
%
}\\
\vspace*{0.5cm}
%
$^{1}$ Instituto de Astrof\'{\i}sica de Andaluc\'{\i}a-CSIC, Glorieta de la Astronom\'{\i}a s/n. P.O. Box 3004 E-18080, Spain\\
$^{2}$ Landessternwarte K\"onigstuhl, Zentrum f\"ur Astronomie der Universit\"at Heidelberg, Heidelberg, Germany\\
$^{3}$ Centro de Astrobiolog\'{\i}a (CSIC-INTA), Madrid, Spain\\
$^{4}$ Institut de Ci\`encies de l'Espai (CSIC-IEEC), Barcelona, Spain\\
$^{5}$ Insitut f\"ur Astrophysik G\"ottingen, G\"ottingen, Germany\\
$^{6}$ Max-Planck-Institut f\"ur Astronomie, Heidelberg, Germany\\
%
\end{flushleft}
%
\markboth{
CARMENES: Terrestrial planets around M dwarfs. A historical overview.
}{ 
%
P.~J. Amado and the CARMENES Consortium
%
}
\thispagestyle{empty}
\vspace*{0.4cm}
\begin{minipage}[l]{0.09\textwidth}
\ 
\end{minipage}
\begin{minipage}[r]{0.9\textwidth}
\vspace{1cm}
\section*{Abstract}{\small
%
CARMENES (Calar Alto high-Resolution search for M dwarfs with Exo-earths with Near-infrared and optical Echelle Spectrographs) is a next generation instrument being built for the 3.5 m telescope at the Calar Alto Observatory by a consortium of eleven Spanish and German institutions. Conducting a five-year exoplanet survey targeting ~300 M dwarfs with the completed instrument is an integral part of the project. The CARMENES instrument consists of two separate \`echelle spectrographs covering the wavelength range from 0.55 to 1.7 $\mu$m at a spectral resolution of R = 82\,000, fed by fibers from the Cassegrain focus of the telescope. The spectrographs are housed in vacuum tanks providing the temperature-stabilized environments necessary to enable a 1 m/s radial velocity precision employing a simultaneous calibration with emission-line lamps.
%
\normalsize}
\end{minipage}
%
%
%
\section{The CARMENES science case \label{science}}
The aim of CARMENES (\cite{Quirrenbach12}) is to perform high-precision measurements of stellar radial velocities with long-term stability. The fundamental science objective is to carry out a survey of late-type main sequence stars (with special focus on moderately active stars of spectral type M4V and later) with the goal of detecting low-mass planets in their habitable zones. For stars later than M4-M5 ($M < 0.20~$M$_\odot$), a radial velocity precision of 1~m/s (per measurement; $\sigma_i$) will permit the detection of super-Earths of 5~M$_\oplus$ and smaller inside the entire width of the habitable zone with 2$\sigma_i$ radial-velocity amplitudes (i.e., $K_{\rm{p}} = 2$~m/s). For a star near the hydrogen-burning limit and a precision of 1~m/s, a planet as small as our own Earth in the habitable zone could be detected. In addition, the habitable zones of all M-type dwarfs can be probed for super-Earths. The CARMENES survey will be carried out with the 3.5-m telescope on Calar Alto, using at least 600 clear nights in the 2014-2018 time frame. We plan to survey a sample of 300 M-type stars for low-mass planet companions (see also \cite{Morales12,Caballero12}). This will provide sufficient statistics to assess the overall distribution of planets around M dwarfs: frequency, masses, and orbital parameters. The seemingly low occurrence of Jovian planets should be confirmed, and the frequency of ice giants and terrestrial planets should be established along with their typical separations, eccentricities, multiplicities, and dynamics.

The project is now at cruise speed towards its final phases of Final Design (FD) and Assembly Integration and Verification (AIV), with most of the funding needed to build the instrument already secured. In this contribution we will also describe how this project was born and the first historical steps taken to consolidate the consortium and reach the current situation. We believe that the details of this process can be of interest to researchers who find themselves at the first stages of an instrumental project in Spain.

\section{CARMENES temporal line \label{Temp}}

\begin{figure}
\vspace{-3mm}
\center
\includegraphics[width=9cm, trim=0cm 8cm 0cm 8cm,clip]{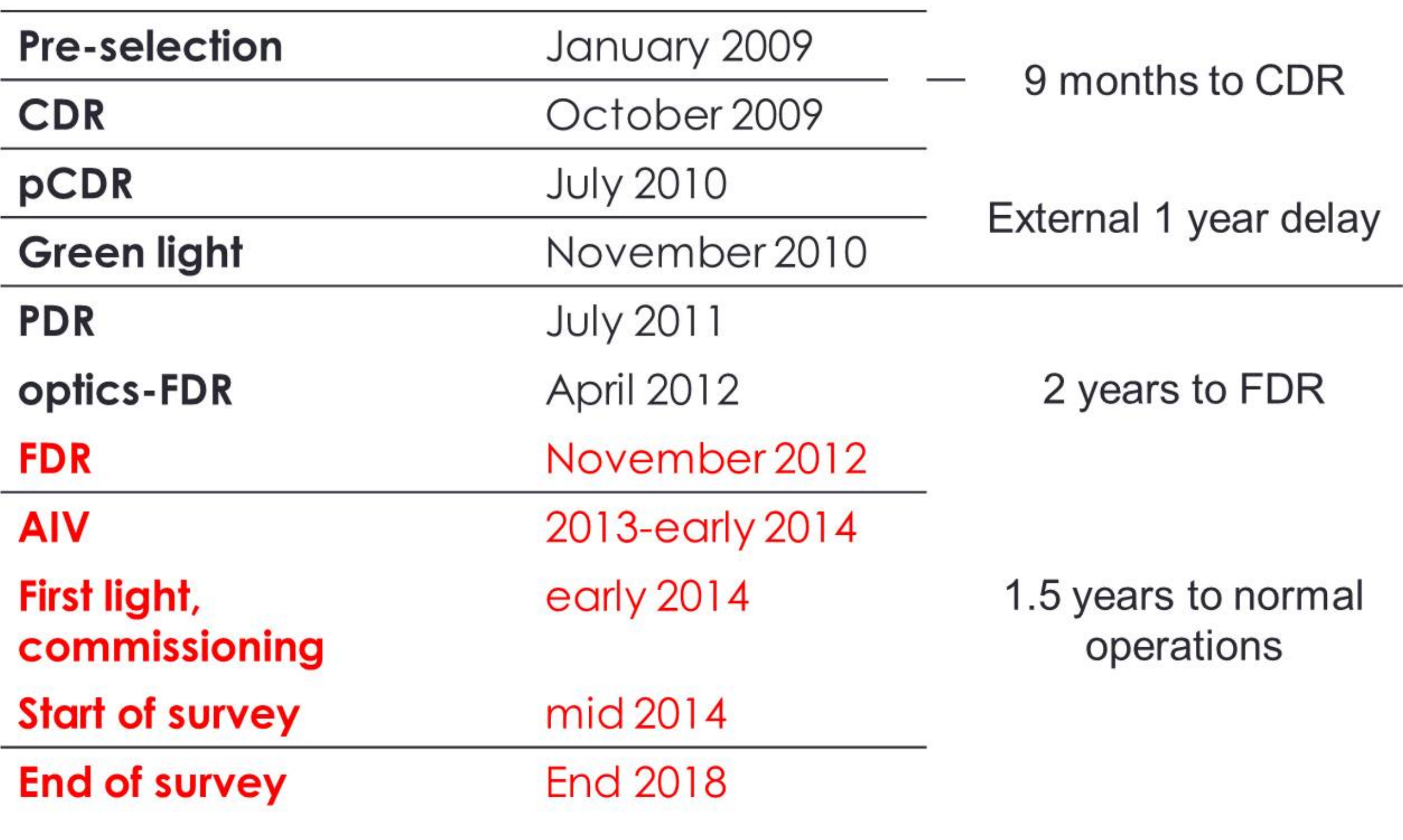} 
\caption{\label{fig1} Temporal line of the CARMENES project. Past phases are shown in black and current and future phases in red.
}
\end{figure}

The time frame for the first stages of the project must be set during 2005, when one of us (PJA) was already part of the Scientific Committee of the project NAHUAL (Principal Investigator (PI) E.~L. Mart\'{\i}n; \cite{Martin10}), to build and exploit a high-resolution, near-infrared (NIR) spectrograph for the Gran Telescopio de Canarias (GTC). The responsibilities in that committee were to represent the interests of the line of research of asteroseismology and magnetic activity in cool stars.

For scientific and economic reasons, E.~L. Mart\'{\i}n, R. Garrido and P.~J. Amado decided, at the end of 2006, to present a similar instrument to a call for proposals for the 3.5-m telescope of the Centro Hispano-Alem\'an de Calar Alto (CAHA) in Almer\'{\i}a, Spain, for which PJA would be the PI. The first step (February 2007) as PI of such a project was to send a ``Letter of Interest'' to the IAA's director, which was immediately forwarded to the CSIC, informing of our intention to design, build and exploit a common-user, fiber-fed, high-resolution spectrograph in the NIR for the 3.5-m telescope at CAHA. The main science case of this project was the search for exoplanets around M-type stars and asteroseismology of pulsating stars with planets. The researchers involved in the project, at that time, were only those mentioned above. The second step was, of course, to raise funds to start supporting the work of our engineers at the Instituto de Astrof\'{\i}sica de Andaluc\'{\i}a (IAA-CSIC) for a conceptual design (CD) and of the researchers at the IAC and the IAA for the preparation of the science. To this aim, we decided to apply for funds to the 6th Framework Programme ERA-Space Technologies Applications and Research for the Regions and medium-sized Countries (ERA-STAR REGIONS) aimed at boosting space research at regions or small countries throughout Europe known to have space research capabilities.

An application to this programme needed of a small consortium to be set up of at least two proposers of two different countries/regions eligible for funding (with space research capabilities). Andaluc\'{\i}a, in the south of Spain, and Madrid both belonged to this program. Therefore, the Universidad Complutense de Madrid (UCM) together with the IAA and two more European nodes, Austria and Lombardy (a region of Italy), were involved in what it would become the first CARMENES (although the project did not have that name at the time) consortium.  Two months later (April 2007) other groups working in different research lines at the IAA were also involved to apply for funds to a programme of the Andalusian Regional Government. Our possibilities of success had seem to be larger at that time, considering that CAHA is sited on Andalusian territory. Eventually, these two proposal did not succeed but engineers and researchers at the IAA kept working on what it was thought to be very interesting project.

The milestone in all this process arrived when the IAA, as co-manager of the CAHA observatory, issued an open call for ideas for new instrumentation for the observatory in March 2008. To this call, seven instruments were presented, among which was CARMENES and other proposals from German institutions based on designs for high-resolution single-object and multi-object spectrographs in the visible wavelength range with similar science cases. PJA came up with the name ``CARMENES'' for the project just the night before his presentation during the workshop that was held at the IAA, a few months later, to manage the results of the call. More information on the reason for this name and the meaning of it can be found at {\tt http://carmenes.caha.es/}.

\begin{figure}
\vspace{-4mm}
\center
\includegraphics[width=9cm]{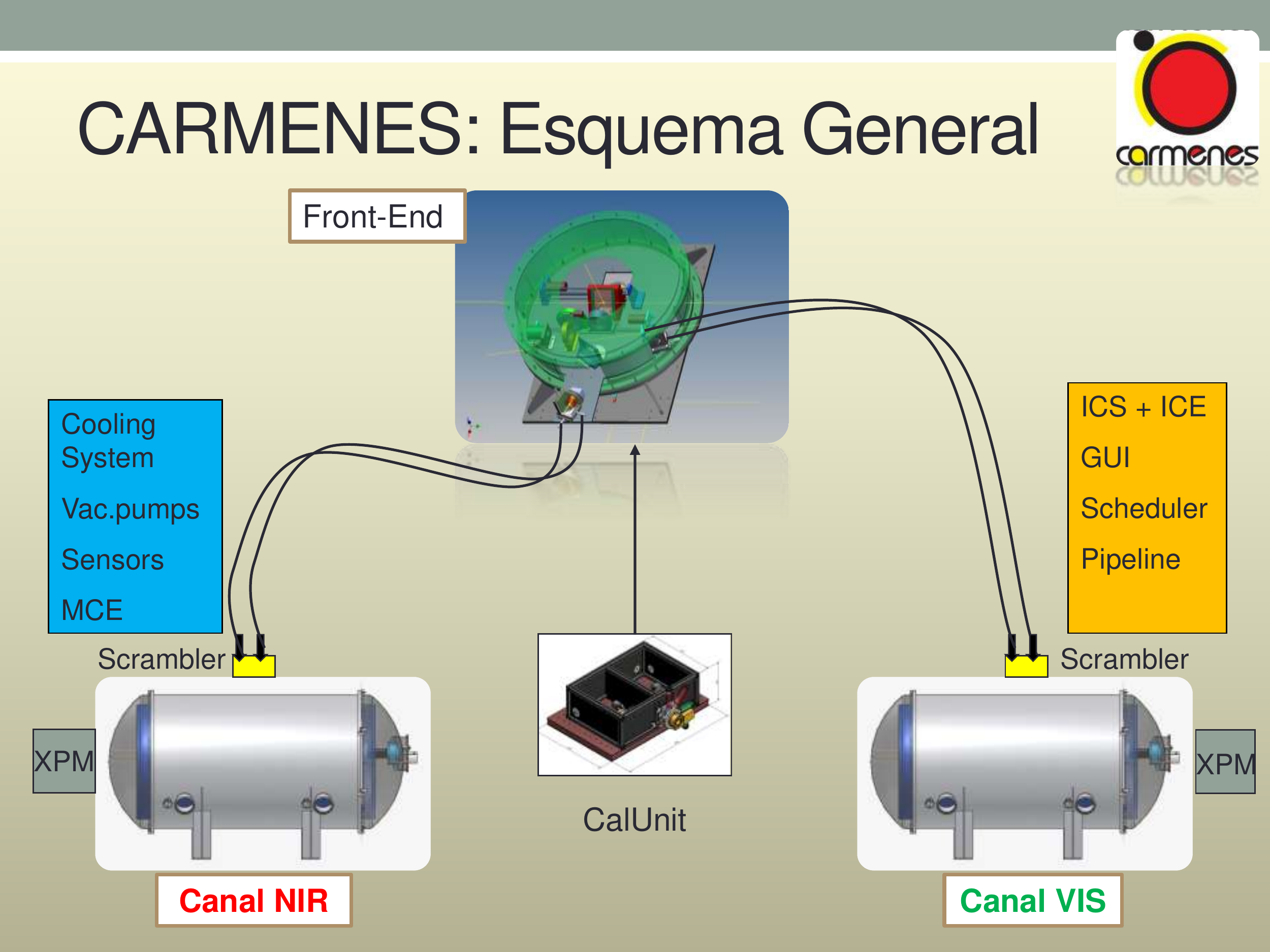} 
\caption{\label{fig2} Layout of the instrument design. Two spectrographs (channels), one in the VIS and one in the NIR wavelength ranges will be thermo-mechanically stabilized in the 3.5-m telescope's Coude room and attached to the telescope with fibres through a front-end adaptor.
}
\end{figure}

From there, things happened very quickly. Contacts with our current German colleagues were initiated from the IAA to suggest a merging of our projects. That increased the probabilities of being selected to be funded for a CD by CAHA. This collaboration started the second era of the CARMENES project. Eventually, a unified proposal for a ``modular'' instrument with three high-resolution channels (spectrographs) was presented in November 2008. The three spectrographs were one single-object and one multi-object, both in the visible and one single-object in the NIR, the latter being the core, non-downscopable part, of the project. The consortium was (is) based on the parity between the Germany and Spain and its structure will be explained below in Sect.~\ref{consortium}.

In January 2009, CARMENES was selected as one out of the two instruments to be funded for a CD by the observatory. To that moment all the instruments built for, and in operation at, CAHA had been proposed by German institutions and, therefore, CARMENES becomes, to date, the first instrument proposed, accepted and co-led by a Spanish institution. During the next few months we developed the CD, which we presented in October 2009 to a CD Review board composed by an international team of experienced researchers and senior engineers. The CAHA Executive Committee, based on the recommendations from this panel, took the decision of selecting CARMENES as the next-generation instrument for the CAHA 3.5-m telescope, culminating with success the work carried out during a period of time of almost three years.

Since then, the project has evolved in a normal way, if for normal we understand the extremely tight time line this ``schedule-driven'' project must keep to have the instrument working at the telescope in 2014. In any case, we went through a ``not-too-normal'' situation during 2009 and 2010 in which Germany and Spain had started negotiations on the agreement for the operation of the observatory for the period 2014-2018, which eventually finished extending the current agreement (2004-2013) for five more years and having CARMENES as the key project for the observatory. In July 2011, we passed successfully a Preliminary Design Review and at the end of this year (2012) we will go through the Final Design Review (FDR). For a brief summary of the CARMENES temporal line see Fig.~\ref{fig1}.

Finally, in 2014 the instrument will be handed to and accepted by CAHA observatory to be used by the Spanish and German communities through both open and guaranteed time. The guaranteed time will sum up to a total of 600 nights in five years (with a possible buffer of another 150 nights), and will be used by the consortium to exploit its main science case. This Guaranteed program is similar to that of the HARPS instrument and similar high-impact results are expected.

\section{The instrument: current status \label{status}}

As mentioned above we are in the final design phase towards the FDR at the end of the year and the Assembly, Integration and Verification (AIV) phase during next year. The CARMENES instrument consists now of two separate \'echelle spectrographs covering the wavelength range from 0.55 to 1.7 $\mu$m at a spectral resolution of $R = 82\,000$, fed by fibers from the Cassegrain focus of the telescope. The spectrographs are housed in vacuum tanks providing the temperature-stabilized environments necessary to enable a 1~m/s radial velocity precision employing a simultaneous calibration with emission-line lamps (see Fig.~\ref{fig2}). Due to strict timeline the project is subjected to, we decided to have a FDR for the optics well in advance of the full instrument FDR to be able to start the procurement of the optical elements, some of which have long-lead times.

The main advantages of this project with respect to similar ones are: 1) the possibility of acquiring simultaneous near-infrared and visible observations, 2) the requirement for both high resolution and wide spectral coverage will be achieved, 3) having long guaranteed time at the 3.5-m telescope for the completion of the project and 4) the project is now leading the development of this type of instrumentation and will, therefore, lead its exploitation and this research topic for the next few years.

\section{The CARMENES consortium \label{consortium}}

\begin{figure}
\vspace{-4mm}
\center
\includegraphics[width=9cm]{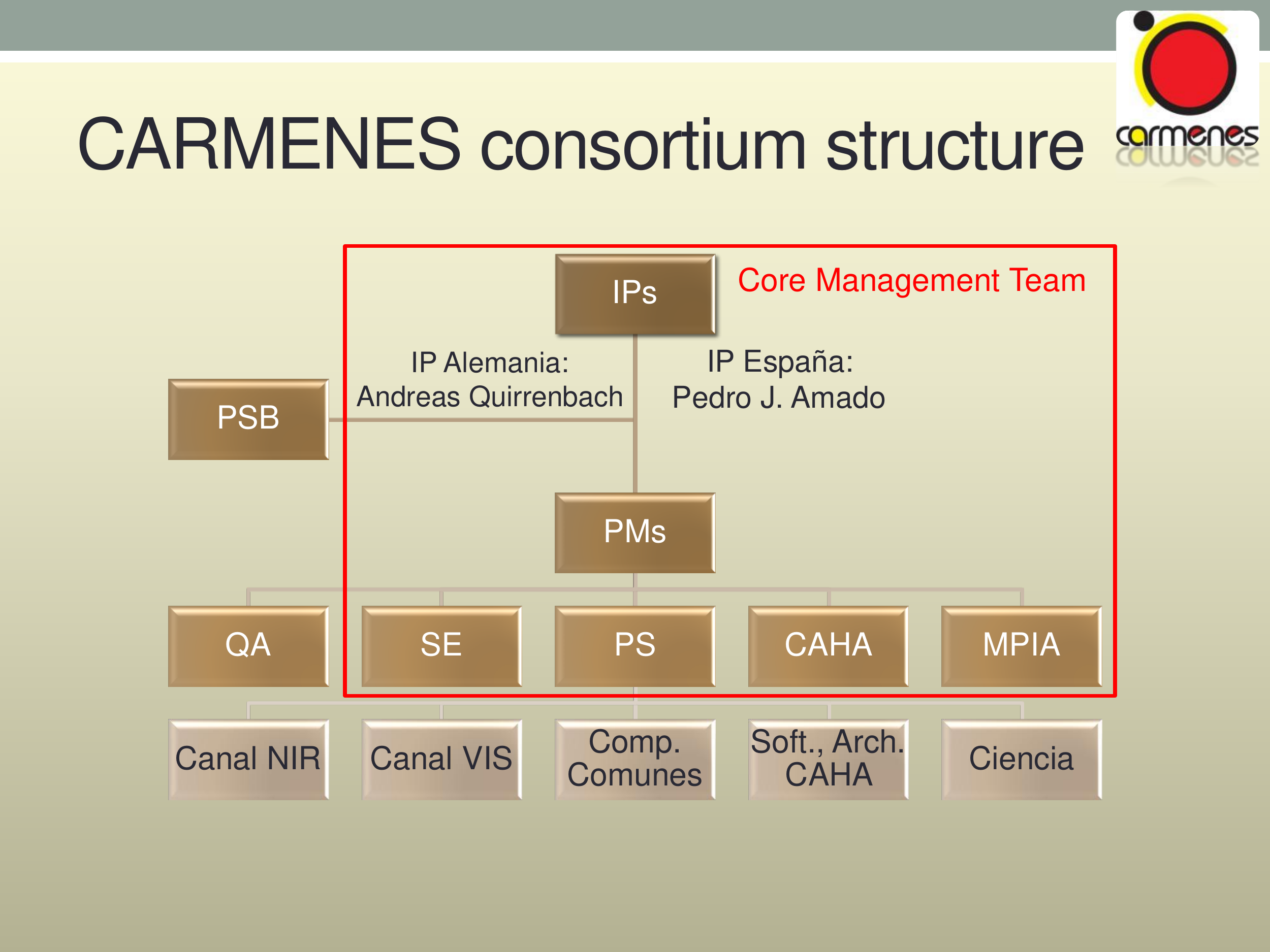} 
\caption{\label{fig3} Schematics of the CARMENES consortium structure. 
}
\end{figure}

The CARMENES consortium is managed on the basis of German-Spanish parity. It was officially created in early 2009 with the aim of designing, building and exploiting the instrument. The consortium is currently made up of eleven centres and universities in Germany and Spain, including the observatory: IAA (Granada), LSW (Heidelberg), MPIA (Heidelberg), CAB (Madrid), HS (Hamburg), IAC (Tenerife), IAG (G\"ottingen), ICE (Barcelona), TLS (Tautenburg) and UCM (Madrid). The member institutions of CARMENES are centres of the German Max-Planck-Gesellschaft and the Spanish Consejo Superior de Investigaciones Cient\'{\i}ficas, as well as some German and Spanish universities and research institutes.

A schematic representation of the consortium structure is given in Fig.~\ref{fig3}. The PIs represent the Spanish and German scientific communities. Two project managers, together with two system engineers, two project sientists, (two posts each, one for a Spanish and one for a German representative), the CAHA representative and the MPIA representative, together with the PIs make up the Core Management Team (CMT). The CMT informs the CARMENES Project Supervisory Board which is composed of the PIs, the directors of the institutions co-managing the observatory (IAA and MPIA) and the observatory's director. To date, there are over 100 researchers and engineers in the consortium.

\section{The CARMENES funding \label{funding}}

The CARMENES instrument will require a total investment of 7.5~MEUR. The cash cost is of around 5.0~MEUR, and the value of manpower contributed in-kind by the members of the CARMENES Consortium is 2.5~MEUR. The amount totaling at least 5~MEUR will be contributed by both CAHA partners, CSIC and MPG, and the partners of the CARMENES Consortium. Most of the funding to be raised by the Spanish partners of the consortium is already available through funded projects from the Ministry of Research and the local Government of the Junta de Andaluc\'{\i}a.

As a sort of conclusion, we can say that, if we consider the ratio between the relatively low cost of this instrument and the highly interesting, competitive and with potentially high-impact-result scientific case, the instrument can be defined as a ``very good instrument''.

%
%
\small  
%
\section*{Acknowledgments}   
%
PJA acknowledges support from the Ministry of Research through grants AYA2010-14840 and AYA2011-30147-C03-01.
%

%
\end{document}